\documentclass[12pt,preprint]{aastex}

\slugcomment{To appear in the Astrophysical Journal}
\shorttitle{The Ionization Fraction of Barnard 68}
\shortauthors{Maret and Bergin}

\newcommand{\eg}{\emph{e.g.}\ }
\newcommand{\ie}{\emph{i.e.}\ }

\begin{document}

\title{The Ionization Fraction of Barnard 68: Implications for Star
  and Planet Formation}

\author{S\'ebastien Maret\altaffilmark{1} and Edwin A. Bergin\altaffilmark{1}}
\affil{Department of Astronomy, University of Michigan, 500 Church
  Street, Ann Arbor, MI 48109-1042, USA}

\begin{abstract}
  We present a detailed study of the ionization fraction of the
  Barnard 68 pre-stellar core, using millimeter
  $\mathrm{H^{13}CO^{+}}$ and $\mathrm{DCO^{+}}$ lines
  observations. These observations are compared to the predictions of
  a radiative transfer model coupled to a chemical network that
  includes depletion on grains and gas phase deuterium
  fractionation. Together with previous observations and modelling of
  CO and isotopologues, our $\mathrm{H^{13}CO^{+}}$ and
  $\mathrm{DCO^{+}}$ observations and modelling allow to place
  constraints on the metal abundance and the cosmic ionization
  rate. The $\mathrm{H^{13}CO^{+}}$ emission is well reproduced for
  metals abundances lower than $3 \times 10^{-9}$ and a standard
  cosmic ray ionization rate. However, the observations are also
  consistent with a complete depletion of metals, \ie with cosmic rays
  as the only source of ionization at visual extinctions greater than
  a few $A_{v}$. The $\mathrm{DCO^{+}}$ emission is found to be
  dependent of the ortho to para H$_{2}$ ratio, and indicates a ratio
  of $\sim 10^{-2}$. The derived ionization fraction is about $5
  \times 10^{-9}$ with respect to H nuclei, which is about an order of
  magnitude lower than the one observed in the L1544 core. The
  corresponding ambipolar diffusion timescale is found to be an order
  of magnitude larger than the free fall timescale at the center of
  the core. The inferred metal abundance suggests that magnetically
  inactive regions (\emph{dead zones}) are present in protostellar
  disks.
\end{abstract}


\section{Introduction}

The ionization fraction (or the electron abundance) plays an important
role in the chemistry and the dynamics of prestellar cores. Because of
the low temperature, the chemistry is dominated by ion neutral
reactions \citep{Herbst73}, and electronic recombination is one of the
major destruction pathways for molecular ions. Furthermore, the
ionization fraction sets the coupling of the gas with the magnetic
field \citep{Shu87}.

Several attempts have been made to estimate the electron fraction in
dense clouds and prestellar cores
\citep{Guelin82,Wooten82,deBoisanger96,Williams98,Caselli98}. These
studies rely on measurements of the degree of deuterium fractionation
(though the $\mathrm{DCO^{+}}$ over $\mathrm{HCO^{+}}$ abundance ratio
for example), which has been found to be roughly inversely
proportional to the electron abundance \citep{Langer85}. However, this
simple approach has caveats \citep{Caselli02d} as it does not consider
line-of-sight variations of the electron fraction. Large density
gradients exist in prestellar cores, and therefore one may anticipate
similar variations in the electron abundance. In addition, the
freeze-out of molecules onto the grain surfaces
\citep[\eg][]{Tafalla02,Bergin02a} influence the degree of deuterium
fractionation independently of the electron fraction
\citep{Caselli98}. Finally, these studies usually consider simple
chemical networks that may neglect important ingredients for the
electron fraction.

In this paper, we study the ionization fraction in the Barnard 68
core, using $\mathrm{H^{13}CO^{+}}$ and $\mathrm{DCO^{+}}$ line
observations. These observations are interpreted with a chemical
network including gas-grain interactions that is coupled to a
radiative transfer model. This technique allows us to infer the
electron abundance along the line-of-sight, and to place constraints
on the abundance of metals, the cosmic ray ionization rate, and the
ionization state of material that is provided by infall to the forming
proto-planetary disk.

The paper is organized as follows: in Section
\ref{sec:observations-results}, we present the observations. The model
used to interpret these observations is detailed in Section
\ref{sec:analysis}. Implications of our findings are discussed in
Section \ref{sec:discussion}, and Section \ref{sec:conclusions}
concludes this paper.

\section{Observations}
\label{sec:observations-results}

The H$^{13}$CO$^{+}$ (1-0) ($\nu = 86.754288$ GHz), and the DCO$^{+}$
(2-1) ($\nu = 144.077319$ GHz) transitions were observed towards B68
($\alpha = 17^\mathrm{h} 22^\mathrm{m} 38.2^\mathrm{s}$ and $\delta =
-23 \degr 49 \arcmin 34.0 \arcsec$; J2000) in April 2002 and September
2002 using the IRAM-30m telescope. The core was mapped with a spatial
sampling of 12\arcsec. The half power beam size of the telescope is
$29\arcsec$ at 87 GHz and $17\arcsec$ at 144 GHz. System temperature
were typically $\sim 110-150$ K at 3 mm and $\sim 160-350$ K at 2
mm. Pointing was regularly checked using planets and was found to be
better than $\sim 2\arcsec$. The data were calibrated in antenna
temperature ($T_\mathrm{a}^{*}$) units using the chopper wheel method,
and were converted to the main beam temperature scale
($T_\mathrm{mb}$), using the telescope efficiencies from the IRAM
website. All observations were carried out in frequency switching
mode. The H$^{13}$CO$^{+}$ (1-0) data were also presented in
\citet{Maret06}.

\begin{figure}
  \centering \includegraphics[scale=.6,angle=-90]{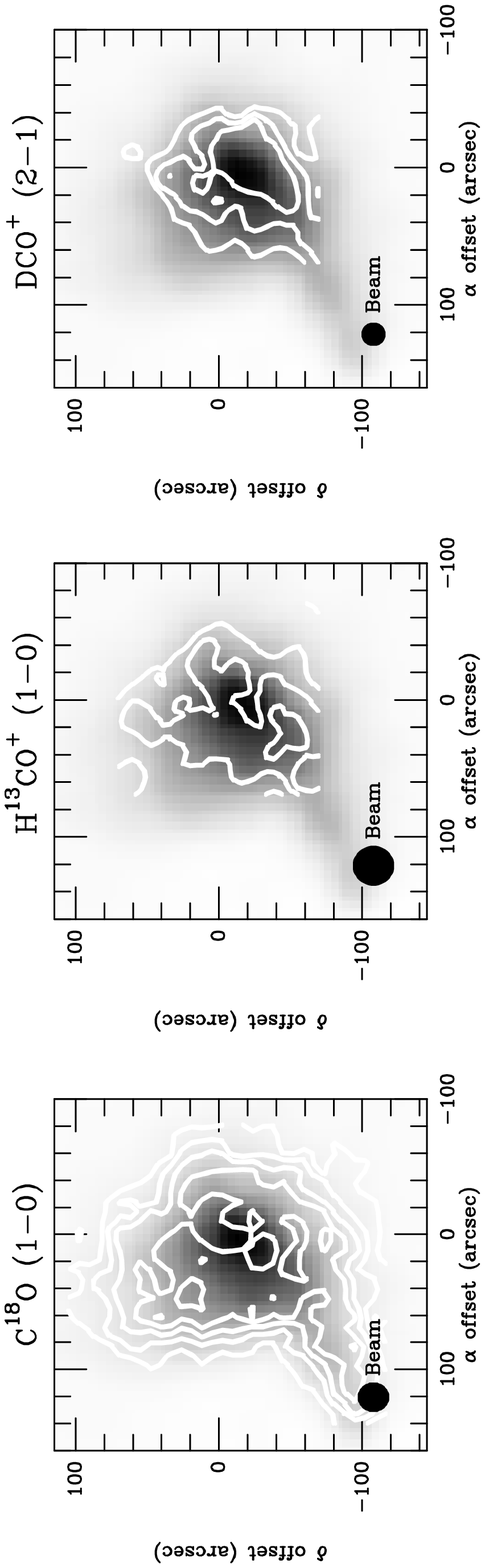}
  \caption{Comparison between integrated intensity maps
    (\emph{contours}) of C$^{18}$O (1-0) (\emph{left}, from
    \citealt{Bergin02a}), H$^{13}$CO$^{+}$ (1-0) (\emph{center}) and
    DCO$^{+}$ (2-1) (\emph{right}) superposed on the map of visual
    extinction obtained by \citet{Alves01}. C$^{18}$O (1-0) contours
    start at 0.2 K km s$^{-1}$ and step by 0.2 K km
    s$^{-1}$. H$^{13}$O (1-0) contours start at 0.15 K km s$^{-1}$ and
    step by 0.15 K km s$^{-1}$. DCO$^{+}$ (1-0) contours start at 0.1
    K km s$^{-1}$ and step by 0.1 K km s$^{-1}$. The $A_{V}$ image
    range from 0 to 27 mag. \label{fig1}}
\end{figure}

Fig. \ref{fig1} shows a comparison between the integrated line
intensity maps of DCO$^{+}$ (2-1) and H$^{13}$CO$^{+}$ (1-0) with the
visual extinction map obtained by \citet{Alves01}. The C$^{18}$O (1-0)
map from \citet{Bergin02a} is also shown.  On this figure, we see that
the peak of H$^{13}$CO$^{+}$ (1-0) line emission does not correspond
to the maximum visual extinction in the core\footnote{This is also
  clearly seen on Fig. \ref{fig2}, which shows the H$^{13}$CO$^{+}$
  (1-0) line emission as a function of the visual extinction. The line
  emission increases as a function of the $A_{v}$ between 0 and 20,
  but decreases at $A_{v} \sim 20$.}. The C$^{18}$O (1-0) line
emission shows a similar behavior: it peaks in a shell-like structure
with a radius of $\sim 50\arcsec$ around the maximum visual
extinction. The DCO$^{+}$ (2-1) line emission, on the other hand,
seems to correlate well with the visual extinction. These differences
are likely a consequence of chemical effects. Because of the
freeze-out on grain mantles, the abundance of CO and its isotopologues
decrease by about two orders of magnitude towards the center of the
core \citep{Bergin02a}. Since H$^{13}$CO$^{+}$ is mainly formed from
the reaction of $^{13}$CO with H$_{3}^{+}$, its abundance is also
expected to decrease towards the core center. DCO$^{+}$ should also be
affected by the depletion of CO. However, the deuterium fractionation
increases as CO is removed from the gas phase. Thus the disappearance
of CO might be compensated by the increased deuterium
fractionation. In the following, we interpret the emission of these
species using a chemical model coupled with Monte-Carlo radiative
transfer model, in order to derive precisely their abundance profiles.

\section{Analysis}
\label{sec:analysis}

We have used a technique that combines the predictions of a chemical
network with a Monte-Carlo radiative transfer
\citep{Bergin02a,Bergin06b,Maret06}. The outline of this technique is
the following. Chemical abundances are computed as a function of the
visual extinction in the core. Using these abundance profiles, the
line emission is computed with a Monte Carlo radiative transfer
code. The resulting map is convolved to the resolution of the
telescope, and is compared to the observations. Free parameters of the
chemical model (\eg cosmic ionization rate, metal abundances, etc.)
are adjusted until a good agreement is obtained between the model and
the observations. Thus, this technique allows for a direct comparison
between the predictions of the chemical network and the observations.

We have used the chemical network of \citet{Bergin95}. This network is
contains about 150 species (including isotopologues, see below), and
focuses on the formation of simple molecules and ions (\eg CO and
HCO$^{+}$). The network includes the effect of depletion on grains,
and the desorption by thermal evaporation, UV photons, and cosmic rays
\citep{Hasegawa93a, Bringa04}. It also includes the effect of
fractionation of $^{13}$C and $^{18}$O, using the formalism described
by \citet{Langer93}. We have extended this network to include the
effect of deuterium fractionation, following the approach used by
\citet{Millar89}. Because of the importance of multiply deuterated
species in the deuterium fractionation process, these species were
also included in the network, following \citet{Roberts04}. It also
include neutralization reactions of ions on negatively charged
grains. The predictions of our network were checked against the UMIST
network \citep{Millar97} for consistency.

We adopt the density profile determined by \citet{Alves01}, from
observations of near infrared extinction from background stars. This
profile is assumed to be constant as a function of time. The dust
temperature profile was computed using the analytical formulae from
\citet{Zucconi01}. For the gas temperature we have adopted the profile
determined by \citet{Bergin06b} from observations and modelling of CO
and its isotopologues. The gas temperature is relatively low (7-8 K),
and increases sightly (10-11 K) at the center of the core as indicated
by ammonia lines observations \citep{Lai03}. This increase in the
temperature is a result of grain coagulation at the center of the
core, which produces a thermal decoupling between the gas and the
cooler dust.

The cloud is supposed to have the initial composition summarized in
Table \ref{table:initial-abundances}. In our model, we assume that the
density profile of the core does not evolve with time. Therefore, we
also assume that the chemistry has already evolved to a point where
hydrogen is fully molecular, and all the carbon is locked into CO.
Our treatment of the initial atomic oxygen pool deserves special
mention. \citet{Bergin02b} examined this question in the context of
the non-detection of water vapor emission in B68 by SWAS. They found
that if atomic oxygen were present in the gas phase in the dense core
center, then the well studied reaction chain that forms
$\mathrm{H_{2}O}$ (via $\mathrm{H_{3}O^{+}}$) would have yielded
detectable water vapor emission. The simplest way to stop this
reaction chain is to remove the fuel for the gas-phase chemistry:
atomic oxygen. This happens when oxygen is trapped on grain surfaces
in the form of water ice \citep[\eg][]{Bergin00}. Thus we have assumed
initial conditions in which all non-refractory oxygen is in the form
of water ice and CO gas with no atomic oxygen left. In this fashion
our initial abundances assume the core formed out of gas that reached
at $A_{v} \sim 2$ -- where $\mathrm{H_{2}}$ and CO have formed and
water ice mantles are observed. On the other hand, nitrogen is assumed
to be mostly in atomic form \citep{Maret06}.

A grain size of 0.1 $\mu m$ is assumed. The cosmic ray ionization rate
and the abundance of low ionization potentials metals ($<$ 13.6 eV)
are free parameters of our study (see \S\ref{sec:metals-depletion} and
\S\ref{sec:constr-cosm-ioniz}). In our models we combine all metals
(\eg Fe$^+$, Mg$^+$, ...) into one species, labeled as M$^{+}$ with
the Fe$^+$ recombination rate of $\alpha(\mathrm{M}^+) = 3.7 \times
10^{-12} (T/300\, \mathrm{K})^{-0.65}\, \mathrm{cm^{3}\, s^{-1}}$.
Due the low ionization potential these metals are assumed to be fully
ionized at the start of the calculation. The network also includes the
neutralization of ions of negatively charged grains with one electron
per grain.
 
The core is assumed to be bathed in a UV field of 0.2 (in Habing
units; \citeyear{Habing68}), as determined by \citet{Bergin06b}. The
chemical abundances are computed as a function of time by solving the
rate equations using the DVODE algorithm \citep{Brown89}. This is done
until a time of $10^5 \ \mathrm{yr}$ is reached. This corresponds to
the ``best-fit'' model of \citet{Bergin06b}. However, as discussed by
\citeauthor{Bergin06b}, this time is a lower limit of the real age of
the cloud, since the CO is assumed to be pre-existing at $t = 0$ in
these models.

Modeling the line emission requires the knowledge of velocity profile
in the core. As a first approach, we have neglected systematic motions
\citep[see][]{Lada03,Redman06}, and we have used the turbulent
velocity profile determined by \citet{Bergin06b} from
$\mathrm{C^{18}O}$ and $\mathrm{^{13}CO}$ lines. The turbulent
velocity is $\sim 0.3$ km/s at the edge of the cloud, and decreases
significantly ($\sim 0.15$ km/s) towards the center of the core.

\section{Results}
\label{sec:results}

\subsection{Metals depletion}
\label{sec:metals-depletion}

Metals ions (\eg Fe$^{+}$ and Mg$^{+}$) play an important role in in
setting the electron abundance in pre-stellar cores, because they are
destroyed relatively slowly by radiative recombination. For example,
the recombination rate of H$_{3}^{+}$ is four order of magnitude
higher than the rate for Fe$^{+}$.

\citet{Guelin82} measured the electron abundance in a sample of dense
molecular clouds using HCO$^{+}$ and DCO$^{+}$ line observations, and
obtained values comprised between 10$^{-8}$ and 10$^{-7}$. The authors
concluded that the metal abundance is lower than 10$^{-7}$ in these
clouds. \citet{Caselli98} determined the electron abundance in a
sample of twenty four low-mass isolated cores (with embedded stars and
starless -- similar in properties to B68) from CO, HCO$^{+}$ and
DCO$^{+}$ observations, and obtained values in the range
10$^{-8}$-10$^{-6}$. \citeauthor{Caselli98} argued that the
differences between cores are due to changes in metal abundance and a
variable cosmic ionization rate ($\zeta$). The best fit between their
chemical model predictions and the observations indicates metal
abundances in the range $2 \times 10^{-9} - 3 \times 10^{-7}$.
\citet{Williams98} determined the electron abundance in a similar
sample of low mass cores using a slightly different approach, and
obtained metal abundances comprised between $5 \times 10^{-9}$ and $4
\times 10^{-8}$ (assuming a constant $\zeta$). All these studies
indicate low metal abundances with respect to their solar
values.. Indeed, observations of FUV FeII absorption lines, and other
metal lines, towards diffuse clouds find depletion factors of over two
order of magnitude with respect to solar values
\citep{Savage79,Jenkins86,Snow02}.

Our $\mathrm{H^{13}CO^{+}} \ (1-0)$ observations can be used to set
limits on the metal ion abundance in B68. $\mathrm{H^{13}CO^{+}}$ is
sensitive to the electron abundance inside the core, because it is
mainly destroyed by electronic recombination. It is also sensitive to
the $\mathrm{H_{3}^{+}}$ and $\mathrm{^{13}CO}$ abundances, since it
is formed from the reaction between these two
species. $\mathrm{H_{3}^{+}}$ itself is mainly formed from
$\mathrm{H_{2}}$ ionization by cosmic rays.  The remaining parameter
in determining the chemical abundance profile is the time dependence
of the chemistry.  In this case, our analysis is simplified because
\citet{Bergin06b} used multiple transitions of $^{13}$CO and C$^{18}$O
and a similar modeling technique to derive the $^{13}$CO abundance and
constrain the ``chemical age\footnote{See discussion in
  \citet{Bergin06b} on the meaning of this ``chemical age''.}'' of
Barnard 68 to $t = 10^5 \, \mathrm{yr}$.  Thus, the only free
parameters for our modeling of the $\mathrm{H^{13}CO^{+}}$ emission
are the cosmic ionization rate $\zeta$ and the metal ion
abundance. These two parameters are difficult to constrain
simultaneously. In \citet{Maret06}, we found that the
$\mathrm{H^{13}CO^{+}} \ (1-0)$ line emission in B68 is well
reproduced by our chemical network if one assume a metal abundance of
$3 \times 10^{-9}$ with respect to H nuclei and a standard cosmic
ionization rate ($\zeta = 3 \times 10^{-17} \ \rm{s}^{-1}$, see next
section). In the following, we explore the parameter space into more
details to place constrains on the metal abundance in the core.

\begin{figure}
  \epsscale{1}
  \plotone{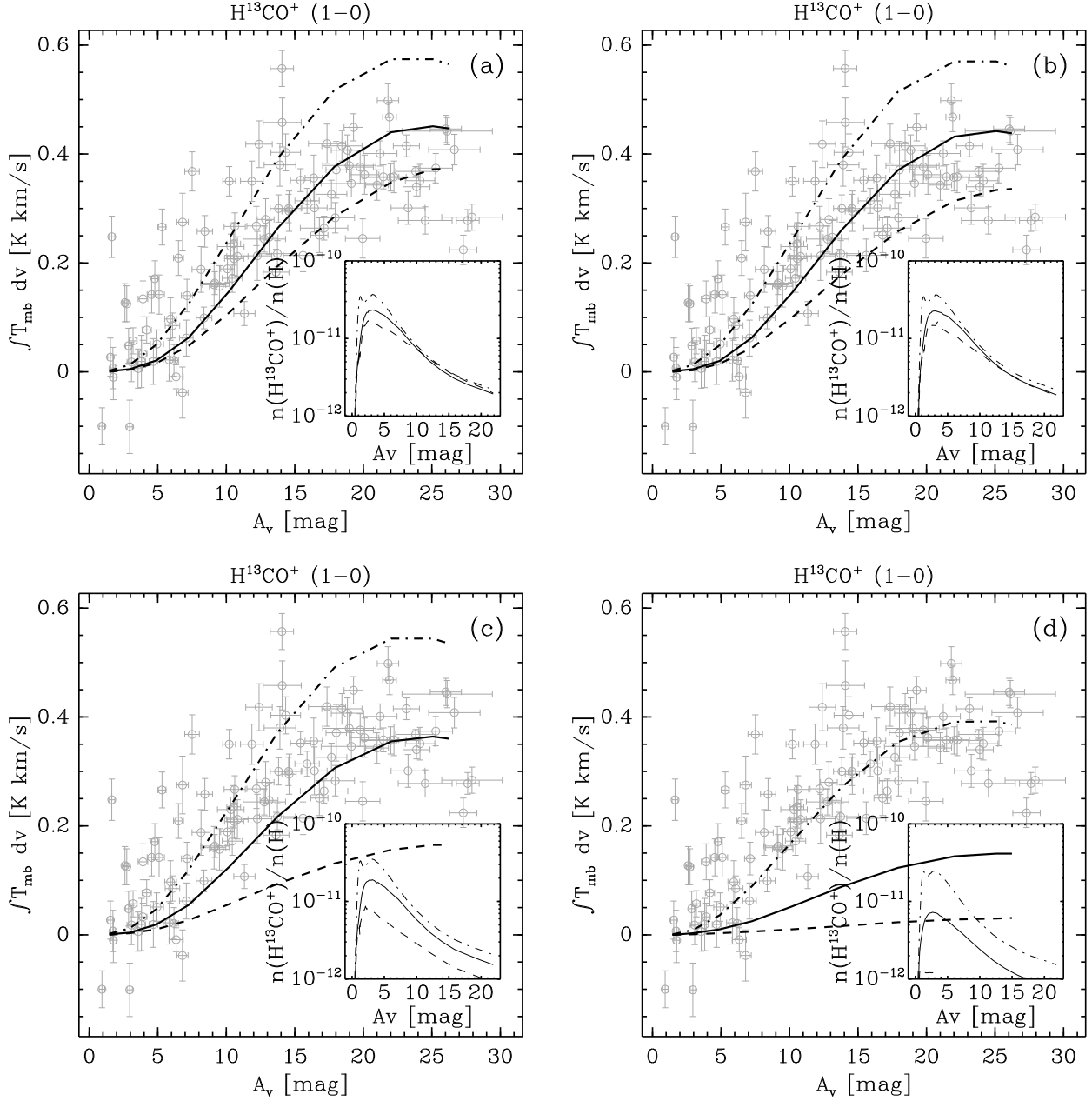}
  \caption{Comparison between the observations and the model
    predictions of the $\mathrm{H^{13}CO^{+}}$ (1-0) line for
    different cosmic ionization rates ($\zeta$) and metal
    abundances. Blue points with error bars (1 $\sigma$) represent the
    observed integrated line intensity as a function of the visual
    extinction in the core ($A_{v}$). Dashed, solid and dash-dotted
    lines represent the model predictions for $\zeta = 3 \times
    10^{-18}$, $3 \times 10^{-17}$ and $3 \times 10^{-16}\, 
    \mathrm{s^{-1}}$ respectively. In panel \emph{(a)} a complete
    depletion of metals is assumed. In panels \emph{(b)}, \emph{(c)},
    and \emph{(d)}, a metals abundance of respectively $3 \times
    10^{-10}$, $3 \times 10^{-9}$ and $3 \times 10^{-8}$ is
    assumed. \label{fig2}}
\end{figure}

On Fig. \ref{fig2}, we show the predicted intensity of the
$\mathrm{H^{13}CO^{+}}$ (1-0) line for different metal ion abundances
and cosmic ionization rates. In these models, metals are assumed to be
initially fully ionized. In Fig. \ref{fig2}, we see that for $\zeta =
3 \times 10^{-17} \ \rm{s}^{-1}$, our model predicts the same
intensities for $x(\mathrm{M}^{+})$ = 0 and $x(\mathrm{M}^{+}) = 3
\times 10^{-10}$. The predicted emission is in fairly good agreement
with the observations.  On the other hand, for a higher metal
abundance ($x(\mathrm{M}^{+}) = 3 \times 10^{-9}$) the model predicts
a intensity slightly lower than the observed, but is in better
agreement with the observations at the center of the core. A metal
abundance of $3 \times 10^{-8}$ is clearly ruled out by the model and
observation comparison. We conclude that $x(\mathrm{M}^{+}) \le 3
\times 10^{-9}$. This value is at the low end of the one obtained by
\citet{Caselli98} and \citet{Williams98}. Compared to the abundance of
metals in the solar photosphere \citep[$x(\mathrm{M}) \sim 8.5 \times
10^{-5 }$;][]{Anders89}, this represent a depletion factor of more
than four orders of magnitude. Indeed, our observations are also fully
consistent with a complete depletion of metals in the core, \ie with
cosmic rays as the only source of ionization at $A_v$ greater than a
few magnitudes (see Fig. \ref{fig2}). It should be noted, however,
that result depends on the value of $\zeta$ adopted. For example, our
observations are fully consistent with a cosmic ionization rate of $3
\times 10^{-16} \ \rm{s}^{-1}$ and $x(\mathrm{M}^{+}) = 3 \times
10^{-8}$. The effects of varying $\zeta$ are discussed in the next
section.

\subsection{Cosmic Ray Ionization Rate}
\label{sec:constr-cosm-ioniz}

Cosmic rays play a crucial role in the chemistry of pre-stellar cores,
because they set the abundance of the pivotal H$_{3}^{+}$ ion, and are
the only source of ionization at $A_{v}$ greater than a few
magnitudes. Despite of its importance, the cosmic ray ionization rate
is difficult to constrain (see \citealt{LePetit04},
\citealt{vanderTak06} and \citealt{Dalgarno06} for recent
reviews). Early estimates in diffuse clouds from HD and OD
observations indicate $\zeta = 7 \times 10^{-17} \, \mathrm{s}^{-1}$
\citep{vanDishoeck86}, a value in agreement with the lower limit of $3
\times 10^{-17} \, \mathrm{s}^{-1}$ measured by the Voyager and
Pioneer satellites \citep{Webber98}. $\mathrm{H_{3}^{+}}$ observations
towards $\zeta$ Persei cloud suggest a significantly higher rate
\citep[$\zeta = 1.2 \times 10^{-15} \, \mathrm{s}^{-1}$;
][]{McCall03}. However, \citet{LePetit04} argued that a value of
$\zeta = 2.5 \times 10^{-16} \, \mathrm{s}^{-1}$ is more consistent
with both $\mathrm{H_{3}^{+}}$ and HD observations. In denser regions,
$\mathrm{HCO^{+}}$ observations indicates a lower ionization rate than
in diffuse clouds: \citet{vanderTak00} obtained $\zeta = (2.6 \pm 1.8)
\times 10^{-17} \, \mathrm{s}^{-1}$ from $\mathrm{HCO^{+}}$ line
observations towards massive protostars. In pre-stellar cores,
\citet{Caselli98} inferred a value comprised between $10^{-18}$ and
$10^{-16} \, \mathrm{s}^{-1}$. The difference in the cosmic ray
ionization rate between diffuse and dense clouds could be due to the
scattering of cosmic rays \citep{Padoan05}. In addition, large
variations are inferred as a function of the Galactic Center distance
\citep{Oka05,vanderTak06}.

Cosmic rays are also heating agents of the gas. \citet{Bergin06b}
examined the value of $\zeta$ in B68 by comparing the predictions of a
chemical and thermal model to observations of CO and its
isotopologues. \citeauthor{Bergin06b} found that their model provide
reasonable fits to the data for $\zeta = 1 - 6 \times 10^{-17} \
\mathrm{s}^{-1}$. Their ``best fit'' model has $\zeta = 1.5 - 3 \times
10^{-17} \ \mathrm{s}^{-1}$. Here we examine the constraints placed by
our $\mathrm{H^{13}CO^{+}}$ observations. On Fig. \ref{fig2}, we see
that our model produces a good fit to the data for $\zeta = 3 \times
10^{-17} \ \mathrm{s}^{-1}$, except for $x(\mathrm{M}^{+}) = 3 \times
10^{-8}$, where the model predictions underestimate the observation by
a factor two. Models with $\zeta = 3 \times 10^{-18} \
\mathrm{s}^{-1}$, consistently underestimate the
observations. Conversely models with $\zeta = 3 \times 10^{-16} \
\mathrm{s}^{-1}$ overestimate the model, except the one with
$x(\mathrm{M}^{+}) \le 3 \times 10^{-8}$. This in agreement with
\citet{Bergin06b}, who found that their observations are not
reproduced by models with $\zeta = 6 \times 10^{-16} \
\mathrm{s}^{-1}$.

To summarize our conclusions regarding the metals abundances and the
cosmic ionization rate, models with $x(\mathrm{M}^{+}) \le 3 \times
10^{-9}$ provide a good agreement with the data\footnote{For
  simplicity, we have assumed that the initial $x(\mathrm{M^{+}})$
  abundance is constant as a function of a radius. A better fit to the
  observations might be obtained with a variation of
  $x(\mathrm{M^{+}})$ with the radius.}, although the model with
$x(\mathrm{M}^{+}) = 3 \times 10^{-8}$ and $\zeta = 3 \times 10^{-16}
\ \mathrm{s}^{-1}$ is also consistent with our data. However values of
$\zeta$ greater that $6 \times 10^{-17} {s}^{-1}$ are ruled out by
\citet{Bergin06b} based on core thermal balance. On the other hand,
models with $\zeta = 3 \times 10^{-18} \ \mathrm{s}^{-1}$ always
underestimate our observations. We conclude that $\zeta = 1-6 \times
10^{-17} \ \mathrm{s}^{-1}$, and $x(\mathrm{M}^{+}) \le 3 \times
10^{-9}$ in B68. This implies that the abundance of ionized metals is
reduced in the center of B68. Charge transfer from molecular ions
(\eg H$_3^+$, HCO$^+$) to metals can be important and a reduction in
the abundance of ionized metals also requires lowering the neutral
metal abundance. In the case of Fe a potential reservoir is FeS
\citep{Keller02}, or organometallic molecules \citep{Serra92}. An
other possility is that Fe is incorporated into grain cores.

\subsection{Ortho to para H$_{2}$ ratio}
\label{sec:ortho-para-h_2}

The ortho to para H$_{2}$ ratio influences the degree of ion and
molecule deuteration in prestellar cores
\citep{PineaudesForets91,Flower06a}. In the gas phase, deuterium
fractionation is mainly due to the following reaction \cite[see ][and
references therein]{Roberts04}:
\begin{equation}
  \mathrm{H_{3}^{+} + HD \rightleftharpoons H_{2}D^{+} + H_{2}}
\end{equation}
The reverse reaction has an activation barrier of $\sim 232 \
\mathrm{K}$ and therefore the reaction becomes essentially
irreversible at low temperature. \citet{Gerlich02} measured the
forward and reverse rates of the above reaction at 10 K, and found
them to be very different than commonly adopted values. The forward
reaction rate was found to be about five times higher than previous
estimates \citep{Sidhu92}, while the reverse reaction rate was found
to be five orders of magnitude larger than previously used
\citep[\eg][]{Caselli98}. In addition, \citet{Gerlich02} determined
via a laboratory measurement that the reverse reaction rate is very
sensitive to the ratio of ortho to para molecular hydrogen. This is
because o-H$_{2}$, in its ground rotational level ($J=1$) has an
higher energy ($\Delta E \sim 170.5 \ \mathrm{K}$) when compared to
the ground state of p-H$_{2}$ ($J=0$). Consequently, o-H$_{2}$ can
more easily cross the energy barrier than p-H$_{2}$, and the rate of
the reverse reaction increases with the ortho to para H$_{2}$ ratio.

Our $\mathrm{DCO^{+}}$ observations can be used to estimate the
$\mathrm{H_{2}D^{+}}$ abundance, and thus the efficiency of the deuterium
fractionation process. $\mathrm{DCO^{+}}$ is mainly formed by the
following reaction:
\begin{equation}
  \mathrm{H_{2}D^{+} + CO \rightarrow DCO^{+} + H_{2}}
\end{equation}
\noindent and is mainly destroyed by electronic recombination. Thus
the $\mathrm{DCO^{+}}$ emission depends on both the CO and
$\mathrm{H_{2}D^{+}}$ abundances, the electron fraction (induced by
cosmic-rays and by pre-existing metal ions), the ortho to para H$_{2}$
ratio, and on time.  Here we benefit from our previous analysis of CO
which constrained the CO abundance and ``chemical age'' and our
analysis of $\mathrm{H^{13}CO^{+}}$ which limit the metal ion
abundance and cosmic ray ionization rate.  Thus the primary free
parameter is the ortho to para $\mathrm{H_{2}}$ ratio ($o/p$) when we
adopt our best fit parameters of $x(M^{+}) = 3 \times 10^{-9}$ and
$\zeta = 3 \times 10^{-17} \, \mathrm{s^{-1}}$.

\begin{figure}
  \epsscale{0.75}
  \plotone{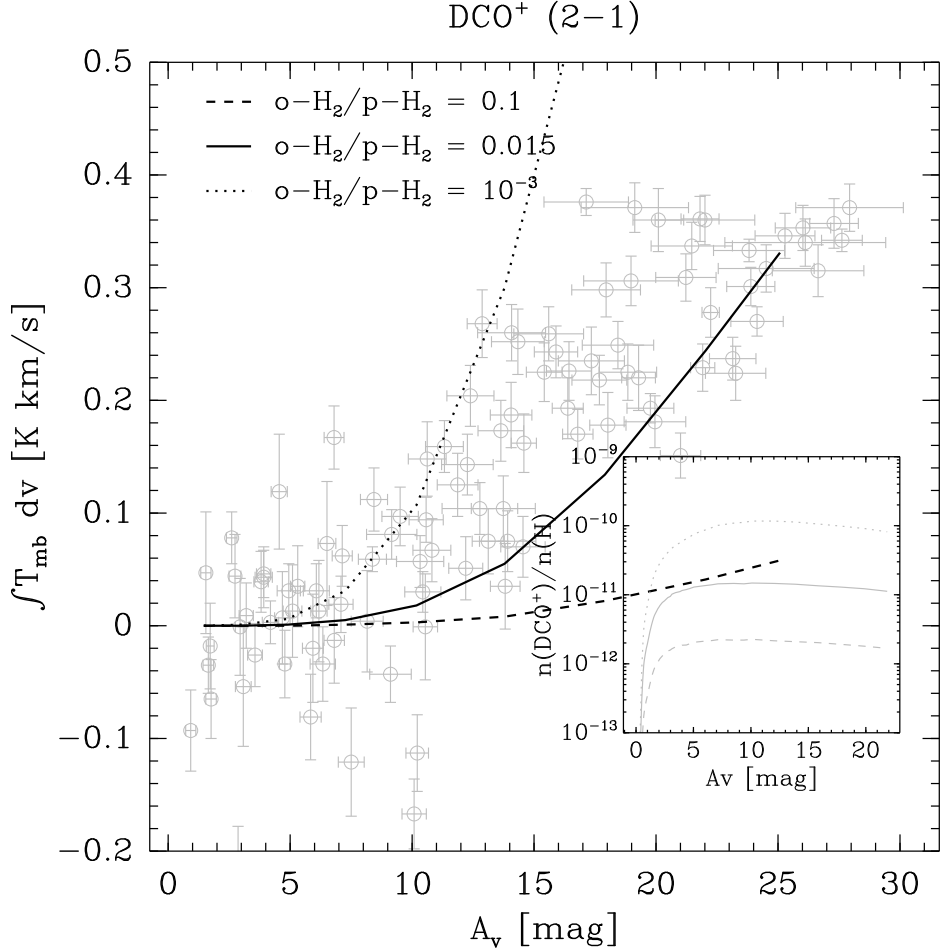}
  \caption{Comparison between the observations and the model
    predictions of the $\mathrm{DCO^{+}}$ (2-1) line for different ortho
    to para $\mathrm{H_{2}}$ ratios. Blue points with error bars (1
    $\sigma$) represent the observed integrated line intensity as a
    function of the visual extinction in the core ($A_{v}$). The dashed,
    dotted and solid black lines show the predicted line intensity for
    different ortho to para $\mathrm{H_{2}}$ ratios. The inset show the
    corresponding abundances of $\mathrm{DCO^{+}}$.\label{fig3}}
\end{figure}

On Fig. \ref{fig3}, we compare the observed $\mathrm{DCO^{+}}$ (1-0)
line emission as a function of $A_{v}$, with the predictions of our
model for different ortho to para H$_{2}$ ratio. Note that in these
models, no $o/p$ conversion is considered: the ortho to para H$_{2}$
ratio is assumed to be constant. The best agreement\footnote{Although
  the model predicts the correct intensity at the core center, one can
  note that emission at lower $A_\mathrm{v}$ is slightly
  underestimated. This may suggest an H$_2$ $o/p$ variation with the
  radius of the core: increasing the ratio at low $A_\mathrm{v}$ would
  increase the $\mathrm{DCO^{+}}$ abundance emission in this region
  and would probably produce a better fit.}  between the observations
and the model is obtained for an $o/p$ ratio of $\sim 1.5 \times
10^{-2}$, well above the Boltzmann equilibrium value at 10 K ($3.5
\times 10^{-7}$). Fig. \ref{fig3} also show the derived
$\mathrm{DCO^{+}}$ abundance inside the core. The abundance peaks at
an $A_{v}$ of $\sim 5$, and decreases slightly towards the core
center, as a consequence of CO depletion (see Section
\ref{sec:electron-abundance}).

It is interesting to compare the $o/p$ H$_{2}$ ratio we obtain with
the predictions of other models. \citet{Walmsley04} modeled the $o/p$
H$_{2}$ ratio in prestellar cores, assuming a complete depletion of
heavy elements. In their model, an initial o/p H$_{2}$ ratio of $3.5
\times 10^{-7}$ is assumed. For a density of $10^6 \
\mathrm{cm}^{-3}$, steady-state is reached in $10^{5}$ yr, a time
comparable to the age of B68 inferred from CO depletion observation
and modeling \citep{Bergin06b}. At steady state, the o/p H$_{2}$ ratio
obtained is $6 \times 10^{-5}$, \ie about two orders of magnitude
lower than the value determined in this work. However, as noted by
\citet{Flower06b}, the ortho to para H$_{2}$ ratio conversion reactions
are very slow, and it is not clear if the steady state equilibrium is
reached in molecular clouds prior to the formation of dense
cores. Using a initial ortho to para ratio of 3 (a value appropriate
for H$_{2}$ formation on grains), \citet{Flower06a} obtain a steady
state ratio of $3 \times 10^{-3}$. This value, although still about a
factor 5 lower, is in better agreement with our estimate. We note that
for $o/p = 3 \times 10^{-3}$, our model predicts an DCO$^{+}$ (1-0)
emission about 2 times higher than the observations (see
Fig. \ref{fig3}).

\section{Discussion}
\label{sec:discussion}

\subsection{Electron abundance and main charge carriers}
\label{sec:electron-abundance}

\begin{figure}
  \epsscale{0.75}
  \plotone{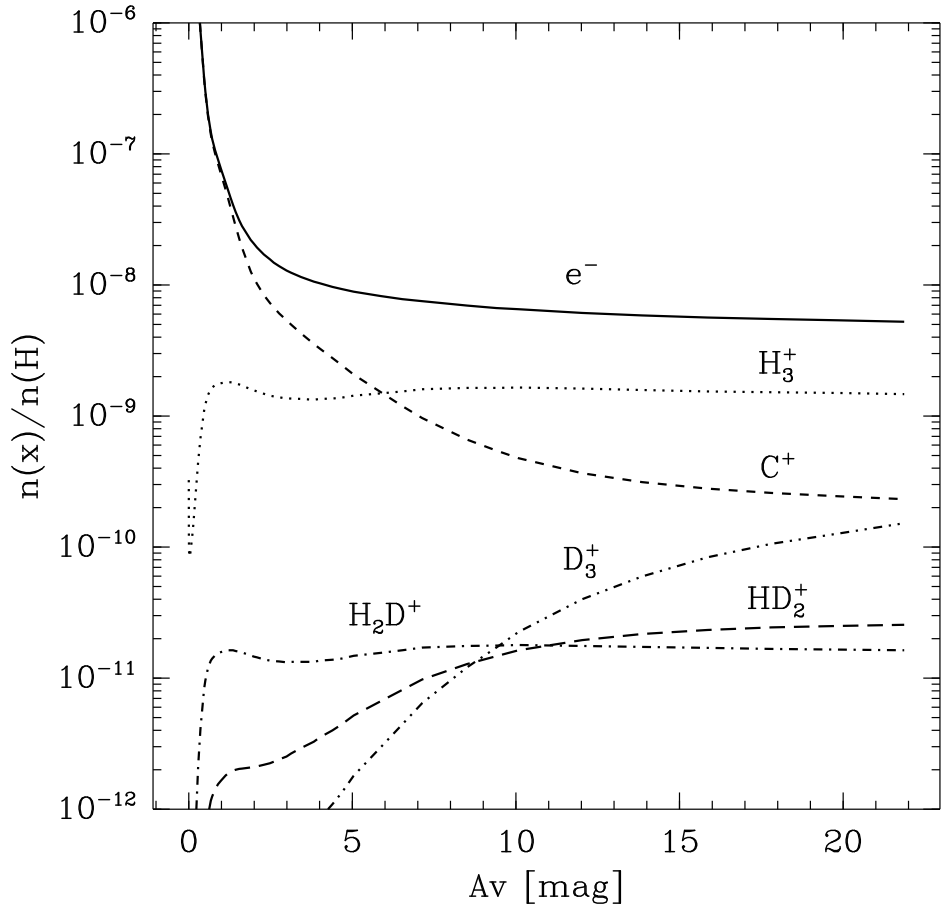}
  \caption{Derived abundances inside B68 for the electrons and main
    ions. Abundances are relative to H nuclei. \label{fig4}}
\end{figure}

On Fig. \ref{fig4}, we show the derived electron and main ions
abundances inside the core. The electron abundance is $\sim 5 \times
10^{-9}$ with respect to H nuclei throughout most part of the core. At
low $A_{v}$, the electron abundance increases as a result of
photo-dissociation of CO. In this region, the most abundant ion is
C$^{+}$. At higher $A_{v}$, the most abundant ion is H$_{3}^{+}$,
which caries about $\sim 20\%$ of the electric charge. The remainder
of the charge is shared between more complex ions. Deuterated ions do
not contribute significantly to the ionization fraction. In the
innermost region of the core, where the deuteration increases as a
result of CO depletion, the main deuterated ion, D$_{3}^{+}$, is about
ten times less abundant than H$_{3}^{+}$. H$_{2}$D$^{+}$ and
D$_{2}$H$^{+}$ have similar abundances ($2 \times 10^{-11}$ with
respect to H). This is in agreement with recent observations
\citep{Vastel04}.

Recently, \citet{Hogerheijde06} reported a probable detection of
$\mathrm{o-H_{2}D^{+}}$ fundamental line towards B68 which can be
compared to our model predictions. The measured flux is however quite
uncertain, given the relatively low signal to noise ratio of this
observation (2.7$\sigma$ and 5.2$\sigma$ on the peak and integrated
intensity, respectively). Assuming a thermal excitation (10 K) and
optically thin conditions, \citeauthor{Hogerheijde06} derive a
$\mathrm{H_{2}D^{+}}$ column density of $1.5 \times 10^{12}$
cm$^{-2}$. Assuming a H$_{2}$ column density of $3.6 \times 10^{22}$
cm$^{-2}$ \citep{Alves01}, this corresponds to an
$\mathrm{H_{2}D^{+}}$ abundance of $2.1 \times 10^{-11}$ with respect
to H nuclei, averaged in the APEX beam (17\arcsec), with respect to H
nuclei. This is in excellent agreement with our model, which predicts
an H$_{2}$D$^{+}$ abundance of $2 \times 10^{-11}$, roughly constant
across the envelope. Of course, if the excitation is non-thermal, the
detection implies an higher abundance. Assuming a 5 K excitation
temperature, \citeauthor{Hogerheijde06} derive a beam averaged
abundance of $1.5 \times 10^{-10}$ with respect to H nuclei. This is
about an order of magnitude higher than our model predictions. Since
no collisional rates exist in the literature for
$\mathrm{H_{2}D^{+}}$, it is unclear whether or not the excitation of
this line is thermal. \citeauthor{Hogerheijde06} estimate a critical
density of $2 \times 10^{6}$ cm$^{-3}$, which exceeds the density at
the center of B68 ($3 \times 10^{5}$ cm$^{-3}$) by about an order of
magnitude. However, the collisional rate, and therefore the critical
density, is uncertain by an order of magnitude
\citep{vanderTak05,Hogerheijde06}. Our model predictions regarding the
deuterium chemistry could be also tested via observations of the
$\mathrm{D_{2}H^{+}} \ 1_{1,0}-1_{0,1}$ ($\nu$ = 691.66044
GHz). Assuming a excitation temperature of 10 K, we predict a line
intensity of 10 mK. Unfortunately, this is too weak to be detected
with current ground based telescopes.

We would like to compare the electron abundance profile we obtained
with the one derived by \citet{Caselli02b} in L1544. In the
\citeauthor{Caselli02b} best fit model, the electron abundance at the
center of L1544 is $5 \times 10^{-10}$ (with respect to H), while we
obtain an electron abundance an order of magnitude higher at the
center of B68. These differences are probably a consequence of
different central densities: the L1544 central density is about an
order of magnitude higher than the one of B68, and the electron
fraction is expected to scale as $n(\mathrm{H}_{2})^{-1/2}$
\citep{McKee89}. Another important difference is the dominant ion:
\citet{Caselli02b} predicts that the most abundant ion is
$\mathrm{H_{3}O^{+}}$, while in our modeling main charge carrier is
$\mathrm{H_{3}^{+}}$. These differences are due to different
assumptions on the atomic oxygen abundance. \citet{Caselli02b} assumes
that oxygen is initially mostly atomic. As a consequence, the
$\mathrm{H_{3}O^{+}}$ abundance is relatively large, because atomic
oxygen reacts with $\mathrm{H_{3}^{+}}$ to form $\mathrm{H_{3}O^{+}}$
(after successive protonations by H$_{2}$ followed by
recombination). In our modeling, oxygen is assumed to be initially
locked in water ices and gas phase CO (see Table
\ref{table:initial-abundances}), and the atomic oxygen gas phase
abundance is relatively low.

Finally, we would like to comment on the effect of grain size
evolution on the electron fraction in the core. \cite{Walmsley04}
computed the electron abundance and main charge carrier in a
prestellar core for different grain sizes. For a grain size of 0.02
$\mu m$, the main charge carrier in their model is H$_{3}^{+}$, while
for larger grains (0.1 $\mu m$), the most abundant ion becomes
H$^{+}$. In their models, H$^{+}$ recombines primarily on grains,
while H$_{3}^{+}$ recombines with free electrons. Since the
recombination timescale on grains depends on the grain size, the
H$^{+}$ over H$_{3}^{+}$ abundance ratio, and in turn the electron
abundance, depends on the grain size as well. However, these models
assume a complete depletion of heavy elements, which is not the case
for B68. In B68 we do find evidence for strong molecular, but not
complete, heavy element freeze-out, at the core center.  The reaction
with H$^{+}$ with molecules containing these elements (\eg NH$_3$,
OH, ...) can transfer the charge to molecular ions with faster
recombination timescales. This would probably reduce the dependence of
the electron abundance on the grain size.

\subsection{Core stability}
\label{sec:core-stability}

The electron abundance in the core is also important for its dynamical
evolution, since its affects the efficiency of ambipolar diffusion. In
a weakly ionized sub-critical core, the ions are supported against
collapse by the magnetic field, but neutrals can slowly drift with
respect to the ions \citep[see ][for a review]{Shu87}. The timescale
for this phenomenon is given by \citet{Walmsley04}:
\begin{equation}
  \tau_{\mathrm{ad}} = \frac{2}{\pi G m_n^{2}} \sum_{i}
  \frac{n_i}{n_n} \frac{m_i m_n}{m_i + m_n} \langle \sigma v \rangle_{in}
\end{equation}
\noindent
where G is the gravitational constant, m$_{n}$ and m$_{i}$ are the
masses of the neutrals and the ions respectively, n$_{n}$ and n$_{i}$
are the number densities, $\langle \sigma v \rangle_{in}$ is the rate
coefficient for the momentum transfer, and the summation goes over all
ions. At low temperature, the rate coefficient for momentum transfer
is \citep{Flower00}:
\begin{equation}
  \langle \sigma v \rangle_{in}= 2 \pi e \left( \alpha \frac{m_i +
      m_n}{m_i m_n} \right)^{1/2}
\end{equation}
\noindent
where $\alpha$ is the polarizability of $\mathrm{H_{2}}$. Assuming
that $\mathrm{H_{3}^{+}}$ is the dominant ion, we obtain:
\begin{equation}
  \tau_{\mathrm{ad}} \sim 2 \times 10^{14} \, x(e)\ \mathrm{yr}
\end{equation}
\noindent
where $x(e)$ is the electron abundance, with respect to H. Thus at the
center at the core, the ambipolar diffusion timescale is $10^{6}$
yr. It is interesting to compare this to the free fall time scale,
which is given by:
\begin{equation}
  \tau_{\mathrm{ff}} = \left( \frac{3 \pi}{32 G \rho} \right)^{1/2}
\end{equation}
\noindent 
where $\rho = n_\mathrm{H_2} m_\mathrm{H_2}$ is the mass
density. When expressed as a function of $n_\mathrm{H_2}$, this gives:
\begin{equation}
  \tau_{\mathrm{ff}} = 3.6 \times 10^{7} \, n_\mathrm{H_2}^{-1/2}\ \mathrm{yr}
\end{equation}
At the center of B68 we obtain $\tau_{\mathrm{ff}} = 7 \times 10^{4} \
\mathrm{yr}$ which is about an order of magnitude faster than the
ambipolar diffusion timescale. Thus, if present, the magnetic field
may provide an important source of support.

The strength of the magnetic field that is needed to support the
cloud can be obtained from the critical mass \citep{Mouschovias76}:
\begin{equation}
  M \sim \frac {0.13}{G^{1/2}} \phi_{B}
\end{equation}
\noindent
where $\phi_{B} = \pi R^{2} B$ is the magnetic flux, $R$ is the core
radius, and $B$ is the magnetic field strength. The strength of the
magnetic field that is needed to support the cloud is therefore:
\begin{equation}
  B \sim \frac {G^{1/2} M}{0.13 \ \pi R^{2}}
\end{equation}
where $M$ is the mass the core. Using $R = 12500\ \mathrm{AU}$ and $M
= 2.1 \ \mathrm{M}_{\sun}$ \citep{Alves01}, we obtain a critical
magnetic field of 76 $\mu$G for B68. No magnetic field measurements
for B68 exist in the literature, but we can compare this value to the
one measured in other cores from dust sub-millimeter
polarization. \citet{WardThompson00} and \cite{Crutcher04} measured
plane-of-the-sky magnetic field strengths of 80 $\mu$G in L183, 140
$\mu$G in L1544 and 160 $\mu$G in L43. \citet{Kirk06} measured lower
fields of 10 and 30 $\mu$G in the L1498 and L1517B cores. Therefore,
if the magnetic field strength in B68 is at the lower end of the
values measured in other cores, then it might be super-critical (\ie
the magnetic field is too weak to balance gravity). If it is higher,
then the core is probably sub-critical.  One may argue B68 has nearly
round shape (albeit with an asymmetrical extension to the southeast),
which potentially is indicative of a weak magnetic field.

\subsection{Implications of the metals depletion for accretion in
  protostellar disks}

One important conclusion of this study is the large metal depletion
inferred for B68. Here we examine the implication of this findings for
the mechanism of angular momentum transport in protostellar disks. The
most favored theory for angular momentum transport in disks predicts
that accretion occurs via magneto-rotational instability \cite[MRI;
][]{Balbus91} which produces MHD turbulence. Since this is a magnetic
process, the ion-neutral coupling is therefore important. Typically,
the ionization fraction should be greater than 10$^{-12}$ for disks to
be able to sustain MHD turbulence \citep[see][and references
therein]{Ilgner06}. \citet{Gammie96} suggested a model wherein the
accretion is layered. The electron abundance is high at the surface of
the disk, because of the ionization of the gas by UV, X-rays and
cosmic rays, but it decreases towards the mid-plane. Thus disks may
have a magnetically active zones at high altitude, where the electron
fraction is sufficient to maintain MHD turbulence, and \emph{dead
  zones}, closer the mid plane of the disk, where the electron
fraction is lower, and accretion cannot occur.

Our results have some import on this process because the chemical
structure of the pre-stellar stage sets the initial chemical
conditions of the gas that feeds the forming proto-planetary disk.
Because of their influence on the ionization fraction, metal ions can
have dramatic effects on the size of the dead zone, assuming that they
are provided by infall to the disk \citep{Fromang02,Ilgner06}. The
latter authors computed the ionization fraction in a protostellar
disk, and found that for $x(\mathrm{M}^{+}) \le 3 \times 10^{-10}$,
the dead zone extend between 0.5 and 2 AU, while it disappears
completely for $x(\mathrm{M}^{+}) \ge 10^{-8}$. In B68, we obtain a
metal abundance of $x(\mathrm{M}^{+}) \le 3 \times 10^{-9}$, which is
below the threshold for a complete disappearance of the dead
zone. Thus, if B68 is representative of the initial conditions for the
formation of protostellar disks, and cosmic rays do not penetrate
deeply to the midplane, dead zones should exist in those disks.

\section{Conclusions}
\label{sec:conclusions}

We have presented a detailed analysis of the electron abundance in the
B68 prestellar core using $\mathrm{H^{13}CO^{+}} (1-0)$ and
$\mathrm{DCO^{+}} (2-1)$ line observations. These observations were
compared to the predictions of time dependent chemical model coupled
with a Monte-Carlo radiative transfer code. This technique allows for
a direct comparison between chemical model predictions and observed
line intensities as a function of radius (or the visual extinction) of
the core. Our main conclusions are:
\begin{enumerate}
\item The metal abundance is difficult to constrain independently from
  the cosmic ionization rate. However, accounting for thermal balance
  considerations and to reproduce $\mathrm{H^{13}CO^{+}} (1-0)$
  emission we estimate that $x(\mathrm{M}^{+}) \le 3 \times 10^{-9}$
  and $\zeta = 1-6 \times 10^{-17} \ \mathrm{s}^{-1}$.
\item The $\mathrm{DCO^{+}} (2-1)$ line emission is sensitive to the
  ortho to para ratio. The emission is well reproduced by our model
  for an ortho to para ratio of $1.5 \times 10^{-2}$, well below the
  equilibrium value, and in reasonable agreement with previous work.
\item The inferred electron abundance is $5 \times 10^{-9}$ (with
  respect to H), and is roughly constant in the core at $A_\mathrm{v}
  > 5$. It increases at lower $A_\mathrm{v}$ because of the
  photo-dissociation of CO and photo-ionization of C. In the dense
  part of the core, the dominant ion is
  H$_{3}^{+}$. $\mathrm{H_{2}D^{+}}$ and $\mathrm{D_{2}H^{+}}$ have
  similar abundances and are about two of magnitude less abundant than
  H$_{3}^{+}$. In the center of the core, our model predicts
  $\mathrm{D_{3}^{+}}$ to be the most abundant deuterated ion.
\item The inferred electron abundance implies an ambipolar diffusion
  timescale of $10^{6}$ yr at the center of the core, which is about
  an order of magnitude higher than the free fall timescale
  ($\tau_{\mathrm{ff}} = 7 \times 10^{4}$ yr).
\item The metal abundance we obtain is below the threshold for
  protostellar disk to be fully active. Consequently, if the chemical
  composition of B68 is reprentative of the initial conditions for the
  formation of a disk and cosmic rays do not penetrate to the disk
  mid-plane, then dead zones should exist in protostellar disks.
\end{enumerate}

\acknowledgments Both authors are grateful to C. Lada for a fruitful
collaboration that led to this work and to T. Huard and E. Aguti for
obtaining a portion of these data. We are also grateful to the referee
and to the editor J. Black for useful and constructive
comments. S.~M. wishes to thanks H. Roberts for helping us testing the
predictions of our network for deuterated species, E. Herbst for
useful discussions about chemical reactions rates, E. Roueff and
M. Walmsley for discussions about the ortho to para ratio, L.
Hartmann and F. Heitch for discussions about the dynamics of B68, and
S. Fromang for discussions about MRI in protostellar disks. This work
is supported by the National Science Foundation under grant 0335207.

{\it Facilities:} \facility{IRAM:30m}, \facility{CSO}, \facility{APEX}

\bibliographystyle{apj}
\bibliography{bibliography}

\clearpage

\clearpage

\clearpage

\clearpage

\clearpage

\begin{deluxetable}{l c}
  \tablewidth{0pt}
  \tablecaption{Initial abundances.\label{table:initial-abundances}}
  \tablehead{
    \colhead{Species} & \colhead{Abundance\tablenotemark{a}}
  }
  \startdata
  $\mathrm{H_{2}}$ & 0.5 \\
  $\mathrm{He}$ & 0.14 \\
  $\mathrm{H_{2}O_{ices}}$ & $2.2 \times 10^{-4}$ \\
  $\mathrm{H_{2}^{18}O_{ices}}$ & $4.4\times 10^{-7}$ \\
  $\mathrm{CO}$ & $8.5 \times 10^{-5}$ \\
  $\mathrm{^{13}CO}$ & $9.5 \times 10^{-7}$ \\
  $\mathrm{C^{18}O}$ & $1.7 \times 10^{-7}$ \\
  $\mathrm{N}$ & $1.50 \times 10^{-5}$ \\
  $\mathrm{N_{2}}$ & $2.5\times 10^{-6}$ \\
  $\mathrm{HD}$ & $1.6 \times 10^{-5}$ \\
  $\mathrm{Grains}$ & $10^{-12}$ \\
  \enddata
  \tablenotetext{a}{Relative to H nuclei.}
\end{deluxetable}

\end{document}